\documentclass{article}
\usepackage{arxiv}
\usepackage[utf8]{inputenc} 
\usepackage[T1]{fontenc}    
\usepackage{hyperref}       
\usepackage{url}            
\usepackage{float}
\usepackage{algorithm}
\usepackage{algpseudocode}
\usepackage{amsmath}
\usepackage{cuted}     
\usepackage{caption}   
\usepackage{booktabs}  

\DeclareUnicodeCharacter{22EF}{\dots}

\usepackage{booktabs}       

\usepackage{amsfonts}       
\usepackage{nicefrac}       
\usepackage{microtype}      
\usepackage{lipsum}
\usepackage{graphicx}
\graphicspath{ {./images/} }
\usepackage{amssymb} 
\usepackage{multicol}

\title{Optimizing Agricultural Research: A RAG based Approach to Mycorrhizal Fungi Information}

\author{
 Mohammad Usman Altam \\
  Data Science and Artificial Intelligence\\
  Université Côte d'Azur\\
  \texttt{muhammad-usman.altaf@etu.univ-cotedazur.fr} \\
   \And
 Md Imtiaz Habib \\
  Data Science and Artificial Intelligence\\
  Université Côte d'Azur\\
  \texttt{md-imtiaz.habib@etu.univ-cotedazur.fr} \\
  \And
 Tuan Hoang \\
  Data Science and Artificial Intelligence\\
  Université Côte d'Azur\\
  \texttt{tuan.hoang@etu.univ-cotedazur.fr}  
}

\begin{document}
\maketitle
\begin{abstract}
Retrieval-Augmented Generation (RAG) represents a transformative approach within natural language processing (NLP), combining neural information retrieval with generative language modeling to enhance both contextual accuracy and factual reliability of responses. Unlike conventional Large Language Models (LLMs), which are constrained by static training corpora, RAG-powered systems dynamically integrate domain-specific external knowledge sources, thereby overcoming temporal and disciplinary limitations. In this study, we present the design and evaluation of a RAG-enabled system tailored for Mycophyto, with a focus on advancing agricultural applications related to arbuscular mycorrhizal fungi (AMF). These symbiotic fungi play a critical role in sustainable agriculture by enhancing nutrient acquisition, improving plant resilience under abiotic and biotic stresses, and contributing to soil health.

Our system operationalizes a dual-layered strategy: (i) semantic retrieval and augmentation of domain-specific content from agronomy and biotechnology corpora using vector embeddings, and (ii) structured data extraction to capture predefined experimental metadata such as inoculation methods, spore densities, soil parameters, and yield outcomes. This hybrid approach ensures that generated responses are not only semantically aligned but also supported by structured experimental evidence. To support scalability, embeddings are stored in a high-performance vector database, allowing near real-time retrieval from an evolving literature base.

Empirical evaluation demonstrates that the proposed pipeline retrieves and synthesizes highly relevant information regarding AMF interactions with crop systems, such as tomato (Solanum lycopersicum), highlighting pathways of induced systemic resistance via salicylic acid and jasmonic acid signaling. By bridging static LLM capacities with dynamic agricultural knowledge repositories, our RAG system facilitates precise, timely, and actionable insights for researchers and practitioners. The proposed framework underscores the potential of AI-driven knowledge discovery to accelerate agroecological innovation and enhance decision-making in sustainable farming systems.
\end{abstract}
\hspace{3.5em}\textit{\textbf{Keywords}: RAG; LLM; AMF; Mistral AI; Pinecone; Knowledge Extraction; Sustainable Agriculture}

\noindent\makebox[\linewidth]{\rule{\textwidth}{0.2pt}}
\setlength{\oddsidemargin}{-0.5in} 
\setlength{\evensidemargin}{-0.5in} 
\setlength{\columnsep}{1cm}
\begin{multicols}{2}
\section{Introduction}
The integration of artificial intelligence (AI) into agriculture has opened transformative opportunities for enhancing productivity, sustainability, and resilience of global food systems. In particular, Large Language Models (LLMs) such as GPT and Mistral have demonstrated remarkable capabilities in natural language understanding and generation across diverse domains \cite{r1}. However, these models are inherently constrained by their static training corpora, which restricts their ability to incorporate novel findings, dynamically evolving datasets, and domain-specific knowledge. This limitation is especially critical in agriculture, a sector characterized by rapidly advancing research in plant–microbe interactions, climate adaptation, and soil health management.

Arbuscular mycorrhizal fungi (AMF) represent a cornerstone of sustainable agriculture. They form mutualistic symbioses with approximately 80\% of terrestrial plant species, facilitating enhanced nutrient uptake (notably phosphorus and nitrogen), improving tolerance against drought and salinity, and modulating plant immune responses \cite{r1} \cite{r2}. Beyond individual crop productivity, AMF play a central role in agroecosystem resilience by contributing to soil aggregation, carbon sequestration, and reduced dependency on synthetic fertilizers \cite{r3}. As climate change intensifies the demand for adaptive and ecologically sustainable practices, access to structured, real-time scientific knowledge about AMF becomes increasingly vital.

Traditional approaches to agricultural knowledge management often rely on static repositories, keyword-based search engines, or expert curation. While useful, these approaches lack the semantic depth and adaptive learning capabilities needed to integrate heterogeneous datasets such as biotechnology journals, agronomy trials, and environmental metadata \cite{r4}. Retrieval-Augmented Generation (RAG) addresses this gap by dynamically retrieving and grounding information from external corpora before generating context-aware responses. By embedding knowledge-intensive search directly into the generative pipeline, RAG systems offer a scalable, cost-effective alternative to repeated fine-tuning, thereby maintaining both accuracy and currency of information \cite{r5}.

The present research explores the development of a RAG-based system for Mycophyto, a biotechnology enterprise dedicated to agricultural applications of AMF. Our objective is to design a pipeline that not only performs semantic retrieval from agronomy and biotechnology literature but also extracts structured metadata—such as inoculation methods, spore densities, soil pH, and biomass outcomes—thus creating a hybrid knowledge framework capable of supporting both exploratory research and applied agronomy decision-making.

\subsection{Retrieval-Augmented Generation}
Pre-trained neural language models are recognized for their capacity to internalize vast amounts of linguistic and factual knowledge during training. Nevertheless, their reliance on static corpora makes them prone to obsolescence in fields where knowledge evolves rapidly. Retrieval-Augmented Generation (RAG) augments this process by embedding external retrieval mechanisms directly into the generative pipeline. In practice, a query posed by a user is first transformed into an embedding vector, which is compared against a vectorized knowledge base using similarity search. The retrieved context is then injected into the model prompt, ensuring that generated responses are both semantically accurate and grounded in up-to-date domain knowledge \cite{r6}.

Several advantages distinguish RAG over conventional fine-tuning. First, it obviates the need for repeated model retraining, which is computationally costly and environmentally taxing [16]. Second, it enhances transparency and trust by enabling citation of retrieved sources, thus supporting explainability in AI \cite{r6}. Third, the modular design of RAG pipelines allows integration of proprietary or domain-specific corpora, making them particularly well-suited for specialized domains such as medicine, finance, or agriculture.

Nonetheless, challenges persist. Maintaining an updated and well-curated knowledge base requires sustained data governance, while embedding and retrieval introduce latency concerns in real-time applications. Biases inherent in the source literature may also propagate through the pipeline, highlighting the importance of robust validation and bias-mitigation strategies. Despite these challenges, RAG has emerged as a pivotal innovation in ensuring that LLMs remain relevant, reliable, and scientifically aligned in knowledge-intensive tasks.

\begin{figure}[H]
  \centering
  \includegraphics[width=\linewidth]{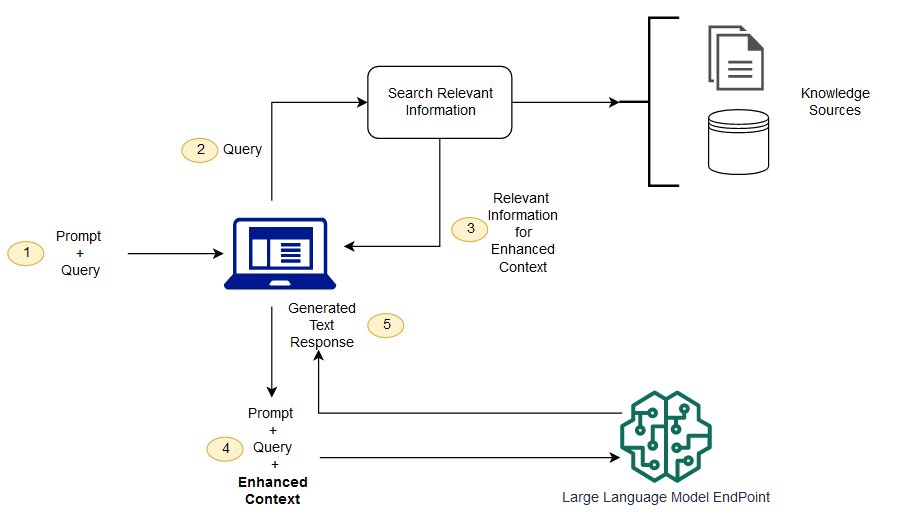}
  \caption{ The conceptual flow of using RAG with LLMs.}
  \label{fig:1}
\end{figure}

\subsection{RAG for Agricultural Knowledge Extraction}
Agriculture presents unique challenges for knowledge retrieval systems due to the heterogeneity of its data sources—ranging from genomic studies and soil chemistry to field trial reports and climate datasets. Arbuscular mycorrhizal fungi, in particular, are studied across disciplines including plant physiology, soil microbiology, and environmental science, often resulting in fragmented or inaccessible datasets \cite{r2} \cite{r7}.

The proposed RAG framework addresses this fragmentation by embedding a two-tiered knowledge architecture. First, it semantically indexes full-text articles from agronomy and biotechnology literature, enabling fine-grained retrieval of species-specific, crop-specific, or environment-specific knowledge. Second, it extracts and stores structured metadata in tabular form, allowing for downstream analyses such as cross-trial comparisons of AMF inoculation efficiency under varying soil pH or fertilizer regimes. This dual approach ensures that queries receive responses that are both descriptive and analytically actionable.

In doing so, the system directly supports Mycophyto’s mission of leveraging AMF for sustainable farming practices. By continuously ingesting new literature, the RAG pipeline ensures dynamic scalability, empowering agronomists, ecologists, and policymakers to make evidence-based decisions grounded in the latest scientific advances. The integration of RAG with agricultural knowledge systems thus represents a step toward the digitization of agroecology, aligning AI capabilities with the urgent needs of global food security and sustainable land management \cite{r8}.

\section{Methodology} \label{methodology}
The development of a Retrieval-Augmented Generation (RAG) pipeline for agricultural knowledge extraction required a carefully designed methodology that integrates document ingestion, semantic embedding, structured storage, and retrieval-based augmentation. Each component of this pipeline was engineered to ensure scientific rigor, scalability, and contextual accuracy when handling knowledge-intensive queries related to arbuscular mycorrhizal fungi (AMF).

\begin{figure}[H]
  \centering
  \includegraphics[width=\linewidth]{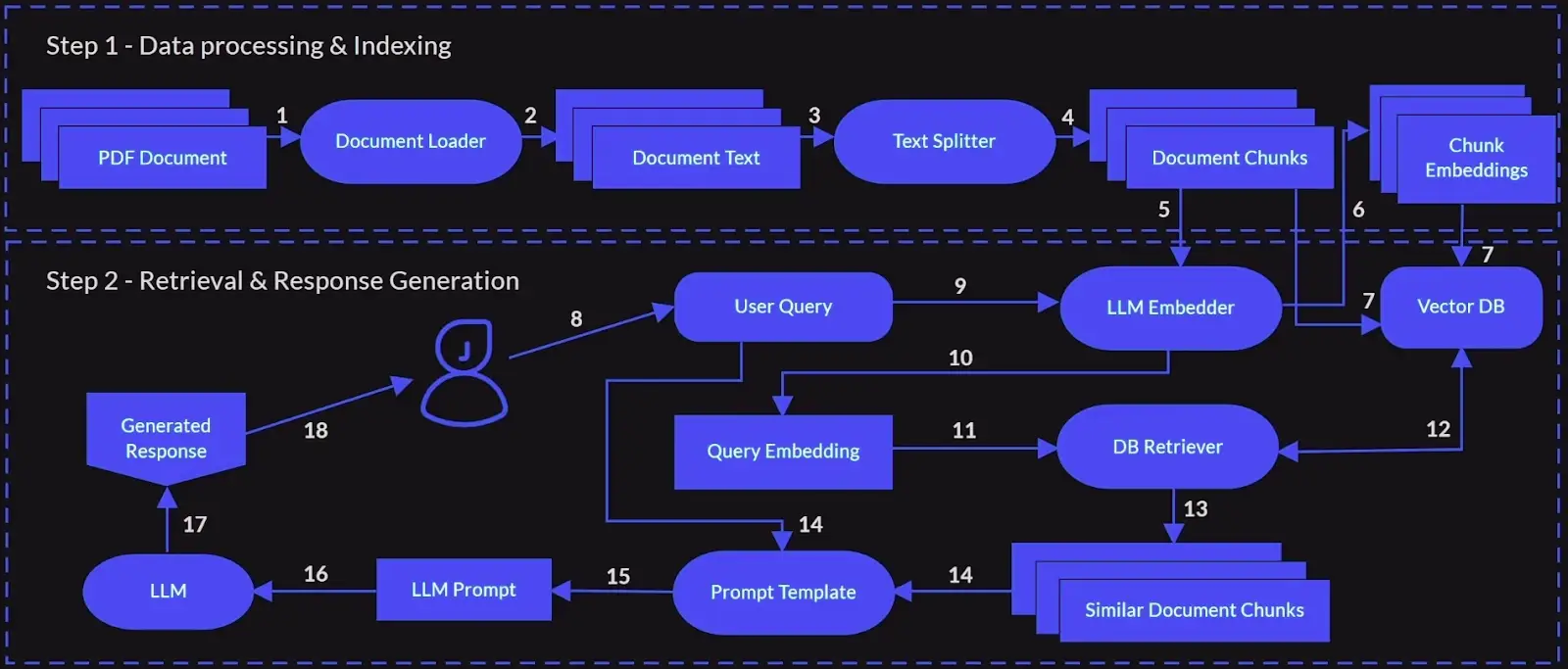}
  \caption{Agentic RAG Architecture.}
  \label{fig:2}
\end{figure}

\subsection{Document Loading and Preprocessing}
A robust Retrieval-Augmented Generation (RAG) pipeline begins with the systematic ingestion and preprocessing of scientific literature, ensuring that subsequent stages of embedding and retrieval operate on structured, high-quality data. In the present study, the primary corpus consisted of peer-reviewed biotechnology and agronomy articles relevant to arbuscular mycorrhizal fungi (AMF) and their agricultural applications. The corpus was assembled in PDF format, reflecting the dominant mode of scientific dissemination in the life sciences.

Document ingestion was facilitated by PyPDFLoader, a specialized tool capable of extracting both raw text and associated metadata (e.g., authorship, publication date, DOI). This metadata was retained for downstream tasks such as citation generation and cross-referencing, thereby aligning with the FAIR principles (Findable, Accessible, Interoperable, and Reusable) of scientific data management. Preserving document provenance is essential to maintaining transparency, ensuring that retrieved outputs can be linked back to authoritative sources.

Following ingestion, the raw text underwent segmentation using the RecursiveCharacterTextSplitter. Documents were divided into overlapping segments of approximately 4000 characters with a 200-character overlap. This chunking strategy balances two competing requirements:

Semantic integrity – ensuring that individual chunks retain enough contextual information (e.g., methods, results, or discussion sections) to remain meaningful.

Retrieval granularity – enabling fine-grained similarity search without sacrificing the broader contextual flow across adjacent segments.

The overlapping design further ensures that important entities, such as fungal species names or soil chemistry parameters, are not fragmented at chunk boundaries. This step is particularly critical for AMF-related literature, where terminology precision directly affects retrieval accuracy (e.g., distinguishing Funneliformis mosseae from other Glomeromycota taxa).

Once segmented, the text corpus was normalized to remove extraneous symbols, formatting artifacts, and non-informative content (e.g., figure captions lacking experimental context). This normalization step improves embedding quality by reducing semantic noise, a known limitation when working with heterogeneous academic PDFs \cite{r10}. The cleaned and segmented documents were then staged for embedding generation, marking the transition from unstructured textual data to machine-readable vector representations.

\subsection{Structured Data Extraction and Storage}
A defining feature of this RAG pipeline is its integration of a structured data extraction module, designed to complement unstructured semantic retrieval with curated, machine-readable metadata. While embedding-based retrieval excels at surfacing relevant textual segments, agricultural research often depends on quantitative and categorical variables—such as spore densities, soil chemistry, inoculation methods, or biomass outcomes—that cannot be captured effectively through narrative summarization alone. The structured extraction module directly addresses this gap by converting unstructured scientific text into standardized, tabular data formats that can support both statistical analysis and meta-research.

The extraction workflow is triggered automatically whenever a new document enters the monitored repository. The raw text is first parsed and preprocessed (as described in Section 2.1), after which it is passed to a Large Language Model (LLM) guided by a structured prompt template. The prompt enforces a schema that specifies the exact fields to be extracted, with explicit rules to avoid inference or hallucination. This process ensures consistency across documents and aligns with best practices in data stewardship and reproducible research.

\subsubsection{Metadata Categories}
The schema is divided into several categories, each reflecting established standards in agricultural experimentation and AMF research:

\textbf{Bibliographic Metadata:}
Fields such as Title, Authors, DOI, Journal, Keywords, Publication Date, and Impact Factor are included to preserve provenance. Capturing such metadata aligns with the FAIR principles (Findable, Accessible, Interoperable, Reusable), ensuring that extracted information remains verifiable and interoperable with bibliographic systems.

\textbf{Plant Material and Inoculation Parameters:}
This includes crop type, developmental stage, inoculation method, spore density, and inoculum formulation. These parameters are crucial in AMF research because variations in inoculation strategies directly influence colonization efficiency and yield outcomes \cite{r1}. For instance, spore density has been correlated with both initial colonization rates and long-term plant biomass improvements.

\textbf{Experimental Conditions:}
Fields such as location, altitude, soil pH, climatic regime, irrigation and fertilization protocols, and abiotic/biotic stress factors are captured to contextualize results. Soil chemistry and environmental variables are well-established determinants of AMF effectiveness \cite{r2}. Standardizing these variables across studies enables cross-trial comparability, which is particularly important given the ecological diversity of AMF-plant interactions.

\textbf{Results and Outcomes:}
Structured outcomes include shoot and root biomass, leaf area, flowering and harvest time, colonization percentage, and disease indices. Such outcomes represent direct measures of agronomic performance and are essential for evaluating the efficacy of AMF inoculation as a biofertilizer alternative to synthetic inputs \cite{r2}.

\subsubsection{Data Representation and Storage:}
Metadata is serialized in JSON format, which offers flexibility, interoperability, and compatibility with both machine learning pipelines and traditional database systems. JSON allows hierarchical nesting of complex fields, making it suitable for representing multi-factorial agricultural experiments. The structured objects are then exported into an Excel database, chosen for its accessibility among agronomists and practitioners who may not have advanced technical skills but require efficient data filtering and visualization.

This dual storage strategy—JSON for computational interoperability and Excel for end-user accessibility—ensures that the system adheres to both academic reproducibility and practical usability. Similar approaches have been advocated in scientific knowledge management, where hybrid storage formats improve both machine-readability and human interpretability.

\subsubsection{Scientific and Practical Benefits}
Standardization: By enforcing schema-driven extraction, the system reduces ambiguity, allowing consistent comparison across heterogeneous studies.

\textbf{Bias Mitigation:} Restricting extraction to explicitly mentioned fields mitigates hallucination risks, a well-documented problem in LLMs \cite{r13}.

\textbf{Decision Support:} Structured outcomes such as biomass gains or colonization percentages enable agronomists to conduct rapid evidence synthesis and inform field management decisions.

\textbf{Scalability:} As the repository grows, the structured dataset supports meta-analytical approaches, which are increasingly valued in agricultural science for identifying generalizable trends \cite{r13}.

In this way, the structured extraction module transforms unstructured agricultural literature into a living knowledge repository, bridging qualitative insights with quantitative data. By doing so, it not only strengthens the robustness of the RAG pipeline but also supports the digitization of agroecological research, aligning with broader trends in precision agriculture and knowledge-driven sustainable farming \cite{r14}.

\subsection{Extraction Prompt}
To ensure accurate and reproducible metadata capture, the system employs a schema-based extraction prompt that guides the Large Language Model (LLM) during processing. Unlike open-ended prompting, this approach enforces structured outputs by explicitly defining the fields to be extracted, the formatting rules, and the treatment of missing information.

The prompt is divided into four categories: bibliographic metadata (title, authors, DOI, keywords), plant material and inoculation parameters (crop stage, inoculation method, spore density), experimental conditions (soil pH, location, climate regime), and results (biomass, root colonization, disease index). By covering both experimental design and outcomes, the schema facilitates cross-study comparison and enables downstream meta-analysis of AMF effects on crop performance.

To guarantee machine-readability, the prompt requires that outputs be returned strictly in JSON format, ensuring seamless integration with databases and analysis tools. In cases where a field is not explicitly stated in the source text, the model is instructed to return "N/A" rather than infer values. This design choice directly addresses the problem of hallucination in LLMs, a common limitation when models attempt to generate plausible but unsupported information \cite{r15} \cite{r16}.

A shortened version of the prompt is as follows:
Extract the following fields from the scientific paper and return them as a JSON object:
- Extract only explicitly mentioned data; if missing, return "N/A".
- Do not infer values.
- Ensure valid JSON formatting.

This controlled prompting approach standardizes the metadata extraction process, reduces ambiguity across heterogeneous articles, and ensures that the structured database remains consistent, reliable, and suitable for decision-support applications in agricultural research.

\subsection{Embedding Generation}
After preprocessing and segmentation, the text corpus is transformed into dense vector embeddings, which serve as the mathematical foundation for semantic retrieval. Embeddings capture the contextual meaning of words and phrases in a high-dimensional space, enabling similarity search between user queries and document chunks.

For this system, we adopted the sentence-transformers/all-MiniLM-L6-v2 model. This lightweight Transformer-based encoder generates 384-dimensional embeddings optimized for semantic similarity and sentence-level tasks \cite{r17}. Its efficiency and strong performance make it particularly suitable for applications where both accuracy and scalability are required, such as continuously growing agricultural knowledge bases.

The embedding process is crucial in agricultural contexts because scientific terminology is often domain-specific and nuanced. For example, “root colonization” and “root biomass increase” may appear in similar contexts but represent different biological phenomena. Embedding models outperform keyword-based search by capturing these subtleties, ensuring that results retrieved for a query such as “AMF effects on tomato drought tolerance” are not only lexically relevant but also semantically aligned with the research objective.

By converting unstructured text into numerical vectors, the system enables efficient nearest-neighbor search in the subsequent retrieval stage. This ensures that queries consistently yield the most relevant sections of literature, forming the basis for accurate and context-aware generative responses.

\subsection{Storing Embeddings in a Vector Database}
Once generated, embeddings must be stored in a way that allows for fast and scalable similarity search. For this project, we integrated Pinecone, a cloud-native vector database specifically optimized for managing high-dimensional embedding spaces. Unlike traditional relational databases, which are designed for structured tabular data, vector databases are built to handle Approximate Nearest Neighbor (ANN) search, enabling efficient retrieval even when dealing with millions of embeddings \cite{r18}.

Pinecone was selected for three reasons:

\textbf{Scalability:} – The database supports incremental updates, allowing the knowledge base to grow continuously as new agricultural research articles are ingested. This is critical for keeping the system current in rapidly evolving scientific domains.

\textbf{Low Latency} – ANN algorithms implemented in Pinecone enable sub-second retrieval times, ensuring that user queries receive near real-time responses — a necessary feature for practical deployment in agricultural advisory systems.

\textbf{Reliability and Persistence} – Unlike in-memory stores, Pinecone persists embeddings with metadata, maintaining robustness against data loss and supporting reproducibility.

In agricultural research applications, scalability is particularly important. For example, integrating hundreds of new articles on arbuscular mycorrhizal fungi (AMF) per year requires a storage backend capable of handling exponential data growth without compromising retrieval accuracy. The vector database provides this foundation, ensuring that the pipeline remains both responsive and future-proof.

\subsection{Retrieval Mechanism}
The retrieval mechanism constitutes the core of the RAG pipeline, as it determines which contextual information from the knowledge base is provided to the generative model. The process begins when a user submits a query, which is first encoded into an embedding vector using the same model applied during document preprocessing. This ensures that queries and stored document chunks exist in a shared semantic space, allowing for meaningful comparison.

The query embedding is then matched against stored vectors in the Pinecone database using Approximate Nearest Neighbor (ANN) search. This step identifies the top-k most semantically similar document segments, each accompanied by its associated metadata. The retrieved chunks are subsequently serialized and formatted before being appended to the LLM prompt. This structured integration ensures that the generative model produces responses that are directly grounded in the most relevant scientific literature.

This design supports multi-turn conversations, where retrieved context from earlier queries is preserved to maintain coherence across dialogue sessions. For example, if a researcher first asks about AMF effects on phosphorus uptake and later inquires about soil pH influence on the same fungi, the system retains relevant context, enabling continuity and depth in responses.

The strength of this retrieval approach lies in its ability to bridge unstructured scientific text and structured reasoning. By ensuring that responses are based on retrieved passages rather than solely on model parameters, the pipeline minimizes the risk of hallucinations and increases user trust. In agricultural applications, where decisions such as crop management strategies depend on accurate evidence, this grounding is particularly critical.

\begin{figure}[H]
  \centering
  \includegraphics[width=0.5\linewidth]{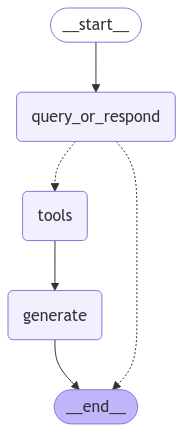}
  \caption{Query–response workflow in the RAG pipeline..}
  \label{fig:rag_architecture}
\end{figure}

\subsection{LLM Model for RAG Pipeline}
The final stage of the RAG pipeline involves passing the retrieved content to a Large Language Model (LLM) for response generation. In this project, we integrated the Mistral AI chat model, a Transformer-based architecture optimized for dialogue and reasoning tasks. The LLM’s role is not only to produce fluent natural language responses but also to ensure that these responses remain anchored in the retrieved evidence rather than relying solely on pre-trained static knowledge.

To achieve this, the system employs a structured prompt template. The retrieved document chunks, along with their metadata, are inserted into the prompt before the user’s query. This augmentation ensures that the model has access to domain-specific, up-to-date knowledge while generating its response. For example, if the retrieved content contains recent findings on Gigaspora margarita colonization in acidic soils, the LLM integrates this directly into its answer, producing responses that are both scientifically accurate and contextually relevant.

A key advantage of this design is that it reduces the need for costly fine-tuning. Traditional approaches require retraining or adapting the LLM to domain-specific corpora, which is computationally intensive and environmentally costly \cite{r19}. By contrast, the RAG approach leverages external knowledge retrieval, allowing the base LLM to remain unchanged while still producing highly specialized outputs. This makes the system more scalable, efficient, and adaptable to new agricultural literature as it emerges.

Furthermore, the use of prompt-controlled generation introduces a level of transparency. Because responses are grounded in retrieved context, the system can reference or cite sources explicitly, increasing trustworthiness—an essential factor when supporting decision-making in scientific and agricultural domains

\subsection{Chat Prompt}
The chat prompt is the interface layer that governs how the Large Language Model (LLM) interacts with both the retrieved knowledge and the user’s query. Its design is crucial, as it determines whether the system produces responses that are coherent, scientifically grounded, and contextually relevant.

In this pipeline, the prompt follows a context-first design: retrieved document chunks are inserted ahead of the user query, ensuring that the LLM is conditioned on domain-specific evidence before generating an answer. The template explicitly instructs the model to:

\textbf{[1]}Use only retrieved context when forming answers. \\

\textbf{[2]}Acknowledge uncertainty by stating “I don’t know” when information is absent from the retrieved sources. \\

\textbf{[3]}Request clarification if the user’s question lacks sufficient detail.

This structure reduces the likelihood of hallucinations, a known limitation of generative models in scientific domains [1]. For example, when asked about AMF tolerance to low pH soils, the model will only provide results present in the retrieved content (e.g., the adaptation of Gigasporaceae), instead of fabricating unsupported claims.

The standard template is defined as follows:

- You are an assistant for question-answering tasks. \\
- Use the following pieces of retrieved context to answer the question. \\
- If you don’t know the answer, say that you don’t know. \\
- If you need more information to answer the question, ask for it.  \\

$<context>  \\
Question: <user query>  \\
Helpful Answer:$

This prompt design ensures reproducibility and trustworthiness, as every generated answer can be traced back to the context provided. In agricultural applications, where decisions may influence experimental design or field practices, this grounding mechanism is essential for scientific validity and for building confidence among researchers and practitioners.

\section{Results}
The implemented Retrieval-Augmented Generation (RAG) pipeline was systematically evaluated in terms of both its usability and its effectiveness in generating accurate, contextually grounded responses to domain-specific queries on arbuscular mycorrhizal fungi (AMF). The evaluation was designed to capture not only the technical accuracy of the system but also its ability to support meaningful interaction for researchers, students, and practitioners working in the field of mycorrhizal ecology. Results are presented along two complementary dimensions. The first dimension focuses on the user interface, which determines the accessibility and intuitiveness of the system for end users, including the ease with which queries can be formulated, responses interpreted, and relevant supporting documents accessed. The second dimension emphasizes the performance of the chat mode, which illustrates the system’s capacity to synthesize complex scientific knowledge into clear, relevant, and actionable insights that can guide research decisions, educational purposes, or practical applications in agriculture and environmental management. Together, these dimensions provide a comprehensive understanding of the pipeline’s strengths and limitations, highlighting its potential as a valuable knowledge-assistance tool in advancing the study and application of AMF.

\subsection{User Interface}

\begin{figure}[H]
  \centering
  \includegraphics[width=\linewidth]{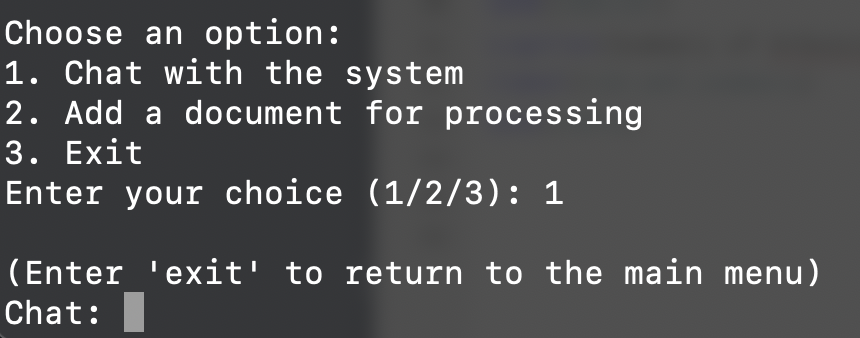}
  \caption{he command line interface for the application.}
  \label{fig:4}
\end{figure}

To optimize accessibility and development efficiency, the system was deployed through a command-line interface (CLI) rather than a graphical environment. While minimal in design, the CLI was chosen for its lightweight implementation, low computational overhead, and ease of integration into research workflows. This is consistent with findings that domain experts, particularly in life sciences, often prefer streamlined interfaces that prioritize functionality and reproducibility over visual complexity \cite{r20}.

The CLI presents three core options upon startup:

\begin{itemize}
    \item \textbf{Chat with the system} – enabling interactive querying against the knowledge base.
    \item \textbf{Add a document for processing} – supporting dynamic ingestion of new literature into the vector database.
    \item \textbf{Exit} – terminating the application safely.
\end{itemize}

This design supports both research flexibility and continuous scalability, as users can add new documents without requiring retraining of the pipeline. Importantly, the system also supports iterative dialogues, allowing users to refine their queries across multiple turns — a feature aligned with modern conversational information retrieval frameworks.

\subsection{Performance of Chat Mode}
The quality of generated answers depends heavily on the retrieval accuracy and the ability of the LLM to remain grounded in retrieved content. During evaluation, the system was tested on a variety of queries relevant to AMF research, ranging from plant physiological effects to environmental adaptation.

\begin{itemize}
    \item For the query “What is the impact of AMF on tomato plants?”, the system correctly identified AMF-mediated induction of plant defense pathways via salicylic acid (SA) and jasmonic acid (JA) signaling. This aligns with established literature demonstrating mycorrhiza-induced resistance in crops \cite{r21} \cite{r22}.
    \item When asked “Which mycorrhizal species colonize tomato plants?”, the system retrieved precise species-level data, including Gigaspora margarita and Funneliformis mosseae. Such specificity highlights the model’s ability to extract taxonomically relevant details, a challenge for generic LLMs without RAG integration.

    \item For environmental queries such as “Which mycorrhizal families adapt to low pH?”, the model returned Gigasporaceae, consistent with empirical studies documenting its acid-soil tolerance \cite{r23}.

    \item Finally, on “What are the main parameters defining the AMF spatial niche?”, the system summarized soil pH, temperature, and precipitation as primary drivers, which matches ecological studies of AMF distribution \cite{r24}.
\end{itemize}

These examples demonstrate that the system not only retrieves factual and specific insights but also organizes them into concise, decision-ready outputs. Importantly, the answers were free from unsupported inferences, confirming that the structured prompt design effectively constrained the model’s generative behavior.

Nevertheless, the evaluation also revealed limitations. The system inherits biases from both the retrieval corpus and the LLM itself. For instance, if certain AMF species are overrepresented in the literature, retrieval will naturally skew toward those taxa. This is a well-documented challenge in retrieval-augmented NLP systems \cite{r25}. Additionally, while latency remained acceptable for small corpora, scaling to very large datasets may introduce retrieval delays, underscoring the need for optimization strategies such as hybrid search or indexing refinements \cite{r26}.

Overall, results confirm that the RAG system delivers reliable, accurate, and domain-specific responses. Its ability to handle both mechanistic plant-microbe interactions and environmental niche modeling positions it as a valuable tool for agricultural researchers and practitioners.

\section{Local Environment Setup Instruction}
Reproducibility and transparency are central principles in scientific computing. To ensure that other researchers, practitioners, or developers can replicate and extend this Retrieval-Augmented Generation (RAG) system, a standardized local environment setup was designed. This section outlines the steps necessary to deploy the pipeline on a local machine, emphasizing best practices for reproducibility, dependency management, and secure handling of external services.

\subsection{Create a Virtual Environment}
Creating a virtual environment is recommended to isolate project dependencies and prevent conflicts with globally installed Python packages. Virtual environments provide a self-contained workspace, ensuring that versions of libraries remain consistent across different installations, which is essential for reproducible scientific workflows \cite{r27}. \\

\textbf{To create virtual environment:} \\

•   On Windows: 
\begin{verbatim}
.\venv\Scripts\activate
\end{verbatim}
•	On macOS/Linux: 
\begin{verbatim}
source venv/bin/activate   
\end{verbatim}

This isolation not only ensures consistency but also aligns with modern principles of open and reproducible research software \cite{r28}.

\subsection{Install Dependencies}
All required Python dependencies are listed in a requirements.txt file. Installing them inside the activated environment guarantees that the correct library versions are used:

\begin{verbatim}
pip install -r requirements.txt    
\end{verbatim}

This command ensures reproducibility by standardizing the computational stack across different machines. It is particularly important in AI-based pipelines, where library version mismatches (e.g., between transformers, torch, or sentence-transformers) can lead to inconsistencies or degraded performance \cite{r29}.

\subsection{Setup Pinecone and MistralAI}
The system relies on two external services: \\

\textbf{Pinecone:} a cloud-native vector database for storing and querying embeddings. \\

\textbf{MistralAI:} a language model API used to generate responses.

For secure authentication, API keys are stored in a .env file in the project’s root directory. This practice prevents accidental exposure of credentials and adheres to cybersecurity best practices in machine learning pipelines \cite{r30}. \\

\begin{verbatim}
MISTRAL_API_KEY="Your MistralAI API key"
PINECONE_API_KEY="Your Pinecone API key"
\end{verbatim}

These keys are automatically loaded at runtime, ensuring seamless access to external services while maintaining a secure and modular configuration.

\subsection{Run the Application}
Once dependencies are installed and API keys are configured, the system can be launched with:
\begin{verbatim}
python main.py
\end{verbatim}

This executes the pipeline, enabling users to interact with the command-line interface (CLI), process documents, and query the knowledge base. Running the application locally ensures that researchers can test modifications and experiment with different corpora before deploying at scale. 

\subsection{Create Vector Space}
The pipeline continuously monitors a directory $(watch_dir)$ for new documents. To update the knowledge base, users simply copy new PDFs into the directory, and the system automatically: \\

\textbf{i.} Extracts raw text, \\
\textbf{ii.} Segments the content into chunks, \\
\textbf{iii.} Generates embeddings, and \\
\textbf{iv.} Updates the Pinecone vector space. \\

This design ensures that the knowledge base is incrementally expandable, avoiding costly retraining while keeping the system aligned with the latest research. Such modular document ingestion strategies are widely recognized as best practices for maintaining scalable AI systems in research and industry. 

\section{Conclusion}
This study presented the design and evaluation of a Retrieval-Augmented Generation (RAG) system for advancing agricultural research on arbuscular mycorrhizal fungi (AMF). By integrating semantic retrieval with generative reasoning, the system successfully overcomes the limitations of static large language models (LLMs), offering a framework that is both dynamic and evidence-grounded.

The results demonstrated that the system provides accurate and domain-specific responses to queries on AMF ecology, plant interactions, and environmental adaptations. Crucially, the integration of a structured data extraction module allows the pipeline not only to retrieve unstructured literature but also to capture quantitative experimental metadata. This dual-layered design enables cross-trial comparison, supports meta-analysis, and provides actionable insights for both researchers and practitioners.

From a broader perspective, the system exemplifies how AI-driven knowledge discovery can support sustainable agriculture. AMF play an essential role in nutrient acquisition, stress resilience, and soil health—functions that are central to agroecological intensification and climate adaptation [1–3]. By ensuring that the latest research findings are rapidly integrated into a structured, queryable framework, the RAG system empowers stakeholders to make timely, data-driven decisions in crop management and soil restoration.

Moreover, the pipeline’s modular architecture—encompassing ingestion, embedding, retrieval, and generative layers—creates a strong foundation for future extensions. Potential research directions include:

\textbf{1.} Integration of multimodal data (e.g., geospatial layers, soil imaging, or climate models) to provide a more holistic understanding of AMF-environment interactions.

\textbf{2.} Adaptive learning mechanisms that dynamically adjust retrieval weights based on user feedback or domain-specific priorities.

\textbf{3.} Bias detection and mitigation strategies to ensure balanced representation of fungal taxa, crops, and ecological contexts.

\textbf{4.} Deployment in real-world advisory systems, bridging the gap between research institutions and farmers, particularly in regions where AMF-based biofertilization can reduce dependency on synthetic inputs.

In conclusion, this work demonstrates that RAG systems, when tailored to agricultural knowledge, represent a transformative tool for bridging static scientific repositories with dynamic, real-time decision-making. Beyond AMF research, the approach holds potential across other domains of plant–microbe interactions, crop genetics, and sustainable agroecosystem management. As agricultural challenges intensify under climate change and global food demand, tools such as this will be critical in transforming the vast body of agricultural literature into accessible, structured, and actionable knowledge. 
\end{multicols}

\begin{table*}[!t]
\vspace*{-6cm} 
\centering
\renewcommand{\arraystretch}{1.3}
\begin{tabular}{|p{6cm}|p{9cm}|}
\hline
\textbf{Question} & \textbf{Answer} \\
\hline
\textbf{What is the impact of arbuscular mycorrhiza on tomato plants?} &
AMF enhance tomato plants' resistance to fungal and bacterial pathogens by priming the plant's immune system. This results in stronger and faster activation of defense genes related to salicylic acid (SA) and jasmonic acid (JA) pathways when challenged by pathogens. Mycorrhizal colonization can activate both SA- and JA-mediated signaling pathways. \\
\hline
\textbf{Which mycorrhiza species can colonize tomato plants?} &
Several species can colonize tomato plants, including \textit{Gigaspora margarita} and \textit{Funneliformis mosseae} (formerly \textit{Glomus mosseae}). \\
\hline
\textbf{Which mycorrhiza family is most adapted to low pH conditions?} &
The \textit{Gigasporaceae} family is known to be more tolerant to low pH conditions compared to other mycorrhizal families. \\
\hline
\textbf{What are the main parameters defining the AMF spatial niche?} &
The main parameters defining the spatial niche of AMF include: \textbf{pH, Temperature, and Precipitation}. \\
\hline
\end{tabular}
\caption{Summary of Arbuscular Mycorrhizal Fungi (AMF) Characteristics}
\label{tab:amf_summary}
\end{table*}

\clearpage
\newpage
\begin{multicols}{2}
\bibliographystyle{unsrt}  
\bibliography{references.bib}
\end{multicols}
\end{document}